\documentclass[twoside,amstex,12pt]{article}
\usepackage{amssymb}
\newlength{\dinwidth}
\newlength{\dinmargin}
\newtheorem{theorem}{Theorem}[section]

\newtheorem{lemma}[theorem]{Lemma}

\newenvironment{definition}{\\[2mm] 
            {\bf Definition:}\it}{\\[2mm]}
\newenvironment{proof}{\medskip \noindent 
            {\bf Proof:}}{\hfill $\square$ \\[2mm] \indent}
\newenvironment{statement}
{\vspace*{-9mm}\begin{list}{}{\setlength{\leftmargin}{2mm}}\item[]}
{\end{list}}
\def\nind{\noindent}

\def\be{\begin{equation}}
\def\ee{\end{equation}}

\def\RR{{\mathbb R}}
\def\CC{{\mathbb C}}

\def\Rd{{\RR^{\, d}}}

\def\bx{{\mbox{\boldmath{$x$}}}}

\def\bp{{\mbox{\boldmath{$p$}}}}
\def\bP{{\mbox{\boldmath{$P$}}}}
\def\bq{{\mbox{\boldmath{$q$}}}}

\def\bP{{\mbox{\boldmath{$P$}}}}
\def\bfp{{\mbox{\footnotesize \boldmath{$p$}}}}
\def\bfx{{\mbox{\footnotesize \boldmath{$x$}}}}

\def\CA{{\cal A}}

\def\CC{{\cal C}}
\def\CH{{\cal H}}
\def\CD{{\cal D}}
\def\CM{{\cal M}}

\def\CO{{\cal O}}

\def\CS{{\cal S}}
\def\CU{{\cal U}}
\def\CV{{\cal V}}
\def\CW{{\cal W}}
\def\CWp{{\CW^{\,\prime}}}

\def\AO{{\CA (\CO)}}

\def\AW{{\CA (\CW)}}

\def\vac{\Omega}
\def\Em{E_m}

\def\pfg{{G}}

\def\inc{{\mbox{\it \scriptsize in}}}
\def\out{{\mbox{\it \scriptsize out}}}

\def\supp{{\mbox{supp}\,}}
\begin{document}
\title{\LARGE Polarization--Free Generators and the S--Matrix}
\author{Hans--J\"urgen Borchers\,$^a$, Detlev Buchholz\,$^a$\, 
and\,Bert Schroer\,$^b$ \\[5mm]
${}^a$ Institut f\"ur Theoretische Physik, Universit\"at G\"ottingen\\
Bunsenstra\ss e 9, 37073 G\"ottingen, Germany\\[2mm]
${}^b$ Fachbereich Physik, Freie Universit\"at Berlin\\
Arnimallee 14, 14195 Berlin, Germany}
\date{\vspace*{5mm} \small Dedicated to the memory of Harry Lehmann}
\maketitle
\begin{abstract}{\noindent   
Polarization--free generators, i.e.\ 
``interacting'' Heisenberg operators which 
are localized in wedge--shaped regions of Minkowski space and 
generate single particle states from the vacuum, are a 
novel tool in the analysis and synthesis of two--dimensional 
integrable quantum field theories. In the present article, the 
status of these generators is analyzed in a general setting. 
It is shown that such operators exist in any theory 
and in any number of spacetime dimensions. But in more than two   
dimensions they have rather delicate domain properties in the 
presence of interaction. If, for example, they are defined and 
temperate on a translation--invariant, dense domain, then the 
underlying theory yields only trivial scattering. In 
two-dimensional theories, these domain properties are consistent 
with non--trivial interaction, but they exclude particle production. 
Thus the range of applications of polarization--free generators  
seems to be limited to the realm of two--dimensional theories.}
\end{abstract}
\section{\hspace{-2mm}Introduction}
\setcounter{equation}{0}
Local quantum field theory provides the adequate 
setting for elementary particle physics. It allows one to 
express in mathematical terms 
the basic features of relativistic quantum physics, 
such as Einstein causality, Poincar\'e covariance and the 
Nahwirkungs--principle, which are encoded in the local 
field equations and commutation relations. The physical interpretation
of the theory relies on asymptotic notions, however, based on the 
particle concept.

The way from the local fields to the asymptotic particle 
interpretation was paved in the seminal work of Lehmann, Symanzik
and Zimmermann \cite{LeSyZi} who invented a consistent collision
theory and established reduction formulas for the computation
of scattering matrix elements. Little is known about the opposite
road, however, i.e.\ the reconstruction of a local theory from
a given scattering matrix. This problem, sometimes called 
form--factor program \cite{KaWe}, is for example 
of importance in the construction of integrable field--theoretic 
models, cf.\ \cite{BaFrKaZa} for some interesting progress 
in this respect.

It was recently noticed \cite{Sch} that certain interacting
theories in two spacetime dimensions admit an important intermediate
step in this program. Namely, there exist  
semi--local polarization--free generators, which are localized 
in wedge--shaped regions of Minkowski space and generate from
the vacuum single particle states, similarly to free fields.
Important features of the theory, such as the crossing symmetry
of the scattering matrix, are encoded in simple analyticity 
properties of the correlation functions of these operators 
(KMS--condition). Moreover, their algebraic properties can directly
be expressed in terms of the elastic scattering amplitudes.
This interesting observation warrants a more systematic investigation
of polarization--free generators in the general setting of
local quantum field theory.

The present article is devoted to such a study. We will show that
there exist polarization--free generators in any local theory with
a non--trivial particle spectrum, irrespectively of the number of
spacetime dimensions. It turns out, however, that these operators
are generically unbounded and their domains of definition exhibit
some delicate features if there is interaction. If, for example, the
polarization--free generators are defined on a translation invariant
domain and the norms of their respective images stay 
polynomially bounded for large translations, then the elastic
scattering amplitudes inevitably vanish in more than two spacetime
dimensions. In two--dimensional theories, these domain properties
are consistent with interaction, but they exclude particle production.
Thus the only non--trivial 
theories in which such temperate families of polarization--free
generators can be defined seem to be models of the type studied in
\cite{Sch}.

The upshot of our investigation is the insight that
polarization--free generators always exist. But in view 
of their subtle domain properties, they are not accessible 
to Fourier analysis in most cases of physical interest. They
are therefore not suitable for a general analysis and synthesis of 
collision states and do not 
provide the desired universal link between the scattering 
matrix and the local interacting fields. Some further aspects 
of our results are mentioned in the conclusions.   

\section{\hspace{-2mm}Existence of polarization--free generators}
\setcounter{equation}{0}
Explicit non--trivial examples of 
polarization--free generators were invented in some 
two-dimensional relativistic quantum field theories 
\cite{Sch}. But their generic 
features can be stated more clearly in the general framework 
of local quantum physics \cite{Ha}. For the convenience of 
the reader who is not familiar with this setting, we briefly
recall in the following the relevant notions and 
explain our notation.  

We assume that we are given a local, relativistic quantum 
field theory in $d$-dimensional Minkowski space $\Rd$. 
But instead of dealing with the unbounded field operators, 
we proceed to their local bounded functions \cite{BoYn}. 
Each of the resulting bounded operators is associated to some  
region $\CO \subset \Rd$, fixed by support of the 
test functions involved in the smearing of the field 
operators. We say that these operators are localized in $\CO$, 
for short. The collection of all operators 
localized in a particular region $\CO$ generates a *--algebra 
$\AO$ on the underlying physical Hilbert space $\CH$ which is closed in the 
weak operator topology. (It is thus a von Neumann algebra.) 
The family of these local algebras inherits from the underlying
quantum field theory the following fundamental properties:

\nind 1.  {(Locality)} The assignment 
\be
\CO \rightarrow \AO 
\ee
defines a net  
over Minkowski space, i.e.\ an inclusion preserving mapping.
The net complies with the principle of locality, that is   
operators affiliated with spacelike separated regions commute. 

Besides these ``local algebras'', we also consider algebras
$\AW$ associated to wedge--shaped regions $\CW$ of the form 
(given in proper coordinates)
\be
\CW_1 = \{ x \in \Rd : x_1 \geq | x_0 |, \ 
x_2, \dots x_{d-1} \ \mbox{arbitrary}\}  \label{2.2}
\ee
as well as their Poincar\'e transforms.
They are the smallest von Neumann algebras containing all
local algebras $\AO$ with $\CO \subset \CW$. Because of locality, 
the algebra $\CA (\CWp)$ associated to the spacelike complement 
$\CWp$ of $\CW$ commutes with $\AW$, $\CA (\CW^{\,\prime}) \subset 
\AW^{\prime}$. 

\nind 2. {(Covariance)} The group of spacetime translations $\Rd$
acts on $\CH$ by a continuous unitary representation $U$
which induces automorphisms of the net. Thus for any
translation $x \in \Rd$ and region $\CO \subset \Rd$
\be
U(x) \AO  U(x)^{-1} = {\cal A} ( \CO + x ) 
\ee
in an obvious notation.

\nind 3. {(Spectrum)} The joint spectrum of the 
generators $P$ of $U$ (the physical energy--momentum spectrum) 
is contained in the closed forward lightcone, 
\be
\mbox{sp}\, P \subset \{p \in \Rd :  p_0  \geq | \bp |  \}.
\ee
Moreover, there is a unit vector $\vac \in \CH$, unique up to a 
phase, which is invariant under the action of
$U$ and cyclic for the local algebras $\AO$ 
(Reeh--Schlieder--property). This vector describes the vacuum state.  

Because of this familiar 
form of the energy--momentum spectrum, the mass operator 
\be
M = (P_0^2 - \bP^2)^{1/2}
\ee
is positive selfadjoint with spectral resolution $E(\,\cdot\,)$.
If there are particles of mass $m$ in the theory,  
the spectral projection  $\Em = E(\{m\})$ is different from zero. It is our
aim to show that there exist operators which are localized in 
wedge regions and generate from the vacuum single particle 
states with mass $m$. 

The formal characterisation of such operators is 
given in the subsequent definition. We recall in this context
that a closed operator is said to be affiliated 
with a von Neumann algebra $\CM$ if it commutes on its domain 
with all elements of the commutant $\CM^\prime$ of  $\CM$. Its 
adjoint is then also affiliated with $\CM$.
\begin{definition}
A closed operator $\pfg$ is called polarization--free 
generator of mass $m$ if (a) it is affiliated with a wedge algebra $\AW$, 
(b) $\vac$ is contained in the domains of $\pfg$ and $\pfg^*$, and
(c) $\pfg \vac$, $\pfg^* \vac$ are elements of $\Em \CH$.
\end{definition}
For the proof that polarization--free generators exist in any
theory, we make use of Tomita--Takesaki--Theory \cite{Ta}.
We begin by recalling some basic facts from this theory for
the case at hand. Since $\vac$ is cyclic and separating for 
the wedge--algebras $\AW$ by the Reeh--Schlieder property and
locality, one can consistently define the Tomita  conjugations
$S_\CW$, setting
\be \label{2.6}
S_\CW A \vac = A^* \vac, \quad A \in \AW.
\ee
These operators are closable anti--linear involutions. Their 
closures, which we denote by the same symbol, have the 
polar decomposition
\be
S_\CW = J_\CW \, \Delta_\CW^{1/2}.
\ee
Here $J_\CW$ is an anti--unitary operator, the modular conjugation,
and $\Delta_\CW$, the modular operator, is strictly positive and 
selfadjoint. The following well--known fact is of fundamental 
importance in the present context. We therefore sketch its proof.
\begin{lemma} Let $\Phi$ be any vector in the domain of 
$S_\CW$. There exists a closed operator $F$ which (a) is affiliated 
with $\AW$, (b) has, together with its adjoint $F^*$, the vector $\vac$
in its domain and (c) satisfies 
\be F \, \vac = \Phi, \quad F^* \, \vac = S_\CW \, \Phi. \ee
\end{lemma}
\begin{proof}
Since the set of vectors $\AW \vac$ is a core fore $S_\CW$ and
$S_\CW$ is closed, there is a sequence 
$F_n \in \AW$ such that $F_n \vac \rightarrow \Phi$ and 
$S_\CW F_n \vac \rightarrow S_\CW \Phi$, strongly. Thus if 
$A^\prime \in \AW^\prime$, one also has 
$F_n A^\prime \vac = A^\prime F_n \vac \rightarrow A^\prime \Phi$ 
and $F_n^* A^\prime \vac = A^\prime F_n^* \vac \rightarrow 
A^\prime S_\CW \Phi$. So the operator $F$, given by
\be
F A^\prime \vac = \lim_{n \rightarrow \infty} 
F_n A^\prime \vac = A^\prime \, \Phi, 
\quad A^\prime \in \AW^\prime,
\ee
is well defined. Its adjoint $F^*$ also has the dense set of vectors
$\AW^\prime \vac$ in its domain and 
\be
F^* A^\prime \vac = \lim_{n \rightarrow \infty} 
F_n^* A^\prime \vac = A^\prime S_\CW \Phi, 
\quad A^\prime \in \AW^\prime.
\ee
This shows that $F$ is closable (we use   
the symbol $F$ also for its closure), and it establishes part
(b) and (c) of the statement since $1 \in \AW^\prime$. For part (a) one 
makes use of the fact that 
for any vector $\Psi^*$ in the domain of $F^*$ and 
$A^\prime, B^\prime \in \AW^\prime$
\be
(A^\prime \Psi^*, F B^\prime \vac) = (\Psi^*, A^{\prime *} B^\prime \Phi) = 
(\Psi^*, F A^{\prime *} B^\prime \vac) =  (A^\prime F^* \Psi^*, B^\prime \vac).
\ee
Hence $|(A^\prime \Psi^*, F B^\prime \vac)| \leq 
\mbox{const} \cdot ||B^\prime \vac ||$,
$B^\prime \in \AW^\prime$. So $A^\prime \Psi^*$ lies in the domain of
$F^*$ and $F^* A^\prime \Psi^* = A^\prime F^* \Psi^*$, 
$A^\prime \in \AW^\prime$. An analogous statement holds for
$F^{**} = F$, so the proof of the lemma is complete.
\end{proof}
In view of this lemma, it suffices for the proof of 
the existence of polarization--free generators
to exhibit non--zero vectors $\Phi_1 \in \Em \CH$
in the domain of $S_\CW$ such that also $S_\CW \Phi_1 \in \Em \CH$.
To accomplish this, we have to take a closer look at the modular operators
and conjugations. Fortunately, we have sufficiently concrete information 
about these objects in the present general setting. 

Since $\Delta_\CW$ is strictly positive, we can
proceed to the corresponding unitary group 
$\Delta_\CW^{is}, s \in \RR$, called modular group.
It is an important consequence of the spectral properties
of the generators of $U$ and covariance \cite{Bo} that
for $x \in \RR^d$
\be
\Delta_\CW^{is} \, U(x) \Delta_\CW^{-is} = U(\Lambda(s) x)
\quad \mbox{and} \quad J^{}_\CW \, U(x) J^{-1}_\CW = U(\Pi x), \label{2.8}
\ee
where $\Lambda(s), s \in \RR$, is (with some appropriate scaling of $s$)
the one--parameter group of boosts leaving the wedge $\CW$
invariant and $\Pi$ is the reflection about the edge of $\CW$. 
If, for example, $\CW_1$ is the wedge given in 
(\ref{2.2}), $x_\pm \in \RR \cdot (\pm 1, 1, 0, \dots 0)$ are any 
two lightlike tranlations in the characteristic planes 
forming the boundary of $\CW_1$, and 
$x_\perp = (0,0,x_2,\dots x_{d-1})$ is any translation
along the edge of $\CW_1$, then 
\be
\Lambda_1(s) x_\pm = e^{\pm 2 \pi s} x_\pm, \ \Lambda_1(s) x_\perp = x_\perp
\quad \mbox{and} \quad \Pi_1 x_\pm =  - x_\pm, \ \Pi_1 x_\perp = x_\perp. 
\ee
Thus the modular groups and conjugations act on the translations
$U$ like Lorentz transformations. As a matter of fact, these 
operators generate a representation of the 
proper Poincar\'e group in generic cases 
according to the Bisognano--Wichmann theorem
\cite{BiWi}. But this more detailed information is not needed here.

Knowing that the modular groups and conjugations act on the 
generators $P$ of $U$ like Lorentz transformations,
we conclude that the mass operator $M$,
being invariant under Lorentz transformations of $P$, 
commutes both with $J_\CW$ and $\Delta_\CW^{is}, s \in \RR$. 
The same is true for the spectral projections $\Em$ of $M$,
hence $S_\CW =  J_\CW \, \Delta_\CW^{1/2}$ 
commutes on its domain $\CD (S_\CW)$ with $\Em$.
This implies that $\Em \CD (S_\CW)$ is a dense subspace of 
$\Em \CH$ which is stable under the action of $S_\CW$
since ${S_\CW}^2 = 1$, cf.\ relation (\ref{2.6}). Applying
the preceding lemma, we have thus established the 
existence of polarization--free generators.
\begin{theorem} 
Given any $m$ in the discrete spectrum of the mass operator 
and any wedge $\CW$, there exist
polarization--free generators $G$ of mass $m$ which are affiliated 
with $\AW$. In fact, for any vector $\Phi_1$ in the dense 
subspace $\Em \CD (S_\CW)$ of $\Em \CH$, there is a $G$ such that 
$ G \vac = \Phi_1$ and $G^* \vac = S_\CW \Phi_1 \in \Em \CD (S_\CW)$.
\end{theorem}

The simplest example illustrating this existence theorem
is free field theory. We briefly discuss it here in order to 
indicate a subtle point in applications of this abstract result.
Let $\phi_0$ be the free massive scalar field acting on Fock 
space. It is well--known \cite{Jo} that 
the field operators $\phi_0 (f)$, smeared with real test functions
$f$ with compact support in some $\CO \subset \Rd$, 
are essentially self--adjoint on the 
domain $\CD_0$ consisting of all vectors with a finite 
particle number. They generate, by their spectral resolutions,
the local algebras $\CA_0 (\CO)$ and are thus affiliated with 
the wedge--algebras $\CA_0 (\CW)$ 
whenever $\supp \, f \subset \CW$. Since the
operators $\phi_0 (f)$ also generate single particle states from the
vacuum, they are polarization free generators in the 
sense defined above.

We mention as an aside that the dense set of vectors $\AW^\prime \vac$
is a core for $\phi_0 (f)$ for any wedge $\CW \supset \supp \, f$.
This implies that, by the preceding general construction,
one would recover $\phi_0 (f)$ from the single particle 
state $\phi_0 (f) \vac$ and the net. 

The full domains of the locally smeared free fields 
$\phi_0 (f)$ are not invariant
under spacetime translations, but they contain the common core 
$\CD_0$ which has this property. As a matter of fact, $\CD_0$
is also invariant under Lorentz transformations and the vector--valued
functions
\be
(\Lambda,x) \rightarrow \phi_0 (f) U_0 (\Lambda,x) \Psi,
\ee
where $U_0$ denotes the underlying unitary representation of the 
Poincar\'e group, are strongly continuous for each $\Psi \in \CD_0$.
Moreover,
\be
|| \phi_0 (f) U_0 (\Lambda,x) \Psi || \leq \mbox{const},
\ee
uniformly for all Poincar\'e transformations $(\Lambda,x)$.

We emphasize that the existence of such a domain $\CD_0$ on 
which polarization--free 
generators exhibit a ``temperate behaviour'' with respect to 
spacetime transformations does not follow from the general   
theorem. But it seems to be an indispensible requirement if one wants
to use these operators in the analysis of collision states and of scattering
amplitudes. For that analysis is based on Fourier transformation, 
which is only meaningful if the underlying functions do not increase
too rapidly at infinity. We therefore take a closer look at such 
temperate generators in the subsequent section.  

\section{\hspace{-2mm}Temperateness and absence of interaction}
\setcounter{equation}{0}
In view of the preceding considerations, we restrict attention 
now to those theories which admit polarization--free generators 
with a temperate behaviour with respect to translations. 
\begin{definition} A polarization--free generator $G$ 
is said to be temperate if there is a dense subspace $\CD$
of its domain which is stable under translations, 
such that for any $\Psi \in \CD$ the function 
$x \rightarrow G  U (x) \Psi$, $\Psi \in \CD$, 
is strongly continuous and polynomially bounded in norm for 
large $x$, and the same holds true also for its adjoint $G^*$. 
The respective subspaces are called domains of temperateness.
\end{definition}
It turns out that this regularity requirement  
imposes severe constraints on the underlying theory 
and excludes interaction if the dimension of spacetime 
is larger than two. In the proof of this statement, 
we restrict attention to massive theories, describing a 
single scalar particle of mass $m$, so the spectrum of $U$ has the form 
\be
\mbox{sp} \, U = \{ 0 \} \cup \{ p \in \Rd : p_0 = (\bp^2 + m^2)^{1/2}\}
\cup \{ p \in \Rd : p_0 \geq (\bp^2 + 4 m^2)^{1/2}\},
\ee
where $m > 0$, but our arguments also apply to 
theories with a more complex particle spectrum.
In a first step we show that temperate polarization--free
generators lead to solutions of the Klein--Gordon equation
and have in their domains single particle states with compact
energy--momentum support about any given point on the ``mass shell''
$\{ p \in \Rd : p_0 = (\bp^2 + m^2)^{1/2}\}$.
\begin{lemma}
Let $G$ be a temperate polarization--free generator of mass
$m$. Then 
\begin{statement}
\item (a) $x \rightarrow G(x) = U(x) G U(x)^{-1}$ is a weak
solution of the Klein--Gordon equation of mass $m$
on the domain of temperateness $\CD$. 
\item (b) The domain $\CD$ contains a dense set of vectors 
with compact spectral support. In particular, there exist 
single particle states of mass $m$ in $\CD$ with spectral 
support in any given neighborhood of any point on the mass shell.
\end{statement}
Corresponding statements hold also for the adjoint $G^*$ of $G$.
\end{lemma}
\begin{proof}
(a) If $G$ is affiliated with the wedge algebra $\AW$, say,  
the operators $G(x)$ 
are affiliated with $\CA(\CW+x)$ by covariance. Now for $x$ varying 
in some open, bounded region $\CU \subset \Rd$, there is a 
wedge $\CW_0 \supset \cup_{x \in \CU} (\CW+x)$, hence the operators
$G(x)^*$, ${x \in \CU}$, contain the common dense subspace 
$\CA(\CW_0)^{\prime} \, \vac$ in their domains. Thus for $\Psi \in \CD$
\be
(G(x)\Psi, A^\prime \vac) = (\Psi, G(x)^* A^\prime \vac)
= (\Psi, A^\prime \ G(x)^* \vac), 
\quad A^\prime \in \CA(\CW_0)^\prime.
\ee
Since $G^*\vac \in \Em \CH$, the function 
$x \rightarrow G(x)^*\vac = U(x) G^* \vac$ 
is a weak solution of the Klein--Gordon equation, so  
one obtains from the preceding equation 
for any test function $f$ with support in $\CU$
\be
\int \, dx \, \big( (\square + m^2) f^* \big) (x) \, 
(G(x)\Psi, A^\prime \vac) = 0, \quad A^\prime \in\CA(\CW_0)^\prime,
\ee
where $f^*$ is the complex conjugate of $f$. 
Making use of the temperateness assumption and the 
Reeh--Schlieder property of $\vac$, this implies
\be
\int \, dx \, \big( (\square + m^2) f\big)(x) \, G(x)\Psi = 0,
\ee
where the integral is defined in the strong sense. Since $\CU$
was arbitrary, the latter equation extends to all test functions
$f \in \CS(\Rd)$ by continuity.\\
(b) The set of vectors of the form $\Psi(f) = \int \! dx \, f(x) U(x) \Psi$, 
where $\Psi \in \CD$ and $f$ is any test function whose Fourier 
transform $\widetilde{f}$ has compact support, 
has compact spectral support and it is dense in $\CH$
since $\CD$ is dense and $U$ is continuous. By choosing the 
support of $\widetilde{f}$ properly, one obtains single particle
states with spectral support in any given neighborhood of any point
on the mass shell. It remains to be shown that these vectors 
belong to the domain of temperateness of $G$. 
There holds for any vector $\Phi^*$ in the domain of $G^*$
\begin{eqnarray}
& |(\Psi(f), G^* \Phi^*)|  = 
| \int \! dx \, f^*(x) \, (U(x) \Psi, G^* \Phi^*)|  
\leq  \int \! dx \, | f(x) | \, | (G U(x) \Psi, \Phi^*) |  &  \nonumber \\
& \leq \int \! dx \, | f(x) | \,  || G U(x) \Psi || \, || \Phi^* || \leq
\int \! dx \, | f(x) | \,  Q(x) \, || \Phi^* || &
\end{eqnarray}
for some polynomial $Q$, depending only on $\Psi$ by the temperateness
assumption. Hence $\Psi(f)$ is an element of the domain of 
$G^{**} = G$, and the same holds true for 
$U(x) \Psi(f) = \Psi(f_x)$, where  
$f_x(y) = f(y-x), y \in \Rd$, is the translated test function. 
The continuity of 
$x \rightarrow G U(x) \Psi(f)$ and temperateness 
follow from the estimate 
$||G\,(U(x) \Psi(f) - \Psi(f))|| \leq 
\int \! dy \, | f_x(y) - f(y)| \, \, Q(y) $  
with the same polynomial $Q$ as above. Hence $\Psi(f)$  
is an element of the domain of temperateness $\CD$. 
The corresponding statements for $G^*$ are established 
in the same manner.
\end{proof}
Picking any single particle state $\Psi_1 \in \CD$ with 
spectral support in a given compact region on the mass shell, 
let us turn next 
to the interpretation of the vectors $G(x) \Psi_1$.
As $x \rightarrow G(x)$ is a solution of the Klein--Gordon
equation, one may expect -- guided by the 
LSZ asymptotic condition -- that 
these vectors describe asymptotic two--particle states. 
But in view of the weak localization properties of the 
generators $G$ and the domain problems involved, some care is 
needed in the analysis.

We rely in our argument on an approach to collision theory 
established by Hepp \cite{He} for the proof of the LSZ reduction formulas in 
the general framework of local quantum field theory, cf.\ also
\cite{Ar}. The main ingredient are quasilocal operators $A(f_t)$ of the form
\be
A(f_t) = \int \! dx \, f_t(x) \, A(x).
\ee
Here $A \in \AO$ are local operators, where the localization 
region $\CO$ is held fixed in the following, and the  
functions $f_t$, $t \in \RR$, are given by 
\be 
f_t (x) = (2\pi)^{-d/2} \! \! \int \! dp \, \widetilde{f}(p)
\, e^{i (p_0 - \omega_p)t} \, e^{-ipx},  \label{3.7}
\ee
where $\widetilde{f} \in \CS (\Rd)$ and 
$\omega_p = (\bp^2 + m^2)^{1/2}$. If $\widetilde{f}$ has support
in a sufficiently small neighborhood of some point on the mass shell, 
$A(f_t) \vac$ is an element of $\Em \CH$ which does not depend 
on $t$. Moreover, 
\be \label{3.8}
\lim_{t \, \rightarrow \, \mp \infty} A(f_t) \, \Phi = 
A(f)_{\!{\inc \atop \out}} \, \Phi, \quad 
\lim_{t \, \rightarrow \, \mp \infty} A(f_t)^* \, \Phi = 
{A(f)_{\!{\inc \atop \out}}}^{\!*} \, \Phi,
\ee
where $A(f)_{\inc}$,  $A(f)_\out$ are the creation operators of an 
incoming, respectively outgoing particle which is in the state 
$A(f) \vac$, and their adjoints 
${A(f)_{\!{\inc}}}^{\!*}$, ${A(f)_{\!{\out}}}^{\!*}$
are the corresponding annihilation
operators. 

These asymptotic relations have been established in \cite{He,Ar} 
for some dense set of ``decent'' collision states $\Phi$. 
But, making use of the fact that  
the operator norms $||A(f_t) E(\Delta)||$ and $||A(f_t)^* E(\Delta)||$
are uniformly bounded in $t$ for any compact subset 
$\Delta$ of the spectrum of $U$ \cite{Bu}, 
they can be extended by continuity to all states with compact 
spectral support.
Thus there holds in particular for the single particle states 
$\Psi_1$ considered above
\be \label{3.9}
\lim_{t \, \rightarrow \, \mp \infty} A(f_t) \, \Psi_1 = 
\big( A(f) \vac \times \Psi_1 \big)_{\!{\inc \atop \out}}, \quad 
\lim_{t \, \rightarrow \mp \, \infty} A(f_t)^* \, \Psi_1 = 
(A(f) \vac, \Psi_1) \, \vac,
\ee
where we employ the standard notation for collision states. 

In the subsequent discussion, we will make use of the
support properties of the functions $f_t$
for asymptotic  $t$ \cite{He}. Let
\be  \label{3.10}
\Gamma(f) = \{\, (1, \bp/ \omega_p) \, 
: \, p \in \supp \widetilde{f} \, \}  
\ee
be the ``velocity support'' of ${f}$
and let $\chi$ be any smooth function which is equal to $1$ on $\Gamma(f)$ 
and vanishes in the complement of some slightly larger region
$\Gamma_\varepsilon(f)$. The asymptotically dominant part 
of $f_t$ is given by  
$x \rightarrow \hat{f_t} (x) = \chi(x/t) f_t(x)$.
It is thus a test function 
with support in $t \, \Gamma_\varepsilon(f)$, $t \neq 0$.
The resulting remainder   
$x \rightarrow \check{f_t} (x) = (1 - \chi(x/t)) f_t(x)$ tends
to zero in the topology of $\CS (\Rd)$ as $t \rightarrow \pm \infty$.
The decomposition $f_t = \hat{f_t} + \check{f_t}$ will be 
repeatedly used in the following arguments. 

Assuming for the sake of concreteness that 
the given generator $G$ is affiliated with 
$\CA (\CW_1)$, where $\CW_1$ is the wedge defined in (\ref{2.2}), 
we introduce the following partial ordering of sets
with reference to that wedge. 
\begin{definition}
Let $\Gamma_a, \Gamma_b \subset \Rd$ be compact sets. 
$\Gamma_a$ is said to be a precursor of $\Gamma_b$, 
$\Gamma_a \prec \Gamma_b$ in formula form, if $\Gamma_b - \Gamma_a$
(the set of all difference vectors) is contained in $\CW_1$.
\end{definition}
Since $\Gamma_a, \Gamma_b$ are compact, 
the set $\Gamma_b - \Gamma_a$ is compact
as well. Hence, as $\CW_1$ is an open cone, 
it follows from $\Gamma_a \prec \Gamma_b$ that 
there is some $\delta > 0$ such that 
$t\, \Gamma_b - t\, \Gamma_a \subset \CW_1 + (0, t \delta,0, \dots 0)$
for $t > 0$. In view of $\{\CW_1 + x\}^{\,\prime} = -\CW_1 + x$, 
this implies $t\, \Gamma_a + (0,t \delta,0, \dots 0) \subset 
(\CW_1 + t \, \Gamma_b)^{\,\prime}$, $t>0$, i.e.\ the
sets $t\, \Gamma_a$  and $(\CW_1 + t\, \Gamma_b)$ are spacelike
separated and their spatial distance increases linearly with $t$.

We shall apply the above order relation to the velocity supports 
of test functions, defined in (\ref{3.10}), as well as to the 
velocity supports of single particle states, which are defined
in an analogous manner by  
\be
\Gamma (\Psi_1) = \{ \,  (1, \bp/ \omega_p) \, : \, p 
\in \supp \Psi_1 \, \}, 
\ee
where $\supp \Psi_1$ is the spectral support of $\Psi_1$.
After these preparations, 
we can clarify now the interpretation of the vectors
$G(g) \Psi_1 = \int \! dx \, g(x) G(x) \Psi_1$ in terms of 
asymptotic two--particle states. 
\begin{lemma} Let $G$ be a temperate polarization--free generator
of mass $m$ which is localized in the wedge $\CW_1$, let $\Psi_1$
be a single particle state in its domain $\CD$ of temperateness 
with compact spectral support, and let $g \in \CS (\Rd)$ be any test
function whose Fourier transform has support in a sufficiently small
neighborhood of some point on the mass shell. Then
\begin{statement}
\item 
(a) \ $ G (g) \Psi_1 = {(G(g) \vac \times \Psi_1 )}_\inc $ \ if \  
$ \Gamma (g) \prec \Gamma ( \Psi_1 ) $, 
\item 
(b) \ $ G(g) \Psi_1 = {(G(g) \vac \times \Psi_1 )}_\out $  \ if \   
$ \Gamma ( \Psi_1 ) \prec \Gamma (g) $.
\end{statement} \label{L2}
\end{lemma}
\begin{proof}
Let $\Psi^*$ be any vector in the domain of temperateness of 
$G^*$ with compact spectral support. Then
\be \label{3.12}
(G(g) \Psi_1, \Psi^*) = \int \! dx \, g^*(x) (\Psi_1, G^*(x) \Psi^*) = 
(\Psi_1, G^*({g}^*) \Psi^*),
\ee
where 
$G^*({g}^*) \Psi^* = \int \! dx \, {g}^*(x) 
G^*(x) \Psi^*$ is defined as a strong integral. Now by the
Reeh--Schlieder property of $\vac$, there is 
for any $\delta > 0$ an $A \in \AO$ and
a test function $f$ whose Fourier transform has support in any 
given neighbourhood of $\supp \Psi_1$ such 
that $||  \Psi_1 - A(f) \vac || < \delta$ and 
$A(f) \vac \in \Em \CH$. In view of the latter fact, one may
replace $f$ in $A(f) \vac$  by any member of the corresponding
family of test functions $f_t$, defined in (\ref{3.7}). Proceeding
to the decomposition $f_t = \hat{f}_t + \check{f}_t$ and 
taking into account that 
$||A(\check{f}_t)|| \leq \int \! dx \, |\check{f}_t (x) | \, || A || 
 \rightarrow 0$ as $t \rightarrow \pm \infty$,
it follows that 
\be 
A(f) \vac = A(f_t) \vac = \lim_{t \,\rightarrow \,\pm \infty}
A(\hat{f}_t) \vac.
\ee
Similarly, since $x \rightarrow G^*(x) \Psi^*$ is a weak solution 
of the Klein--Gordon equation according to Lemma 3.1, one may replace
$g$ in $G^*({g}^*) \Psi^*$ by $g_t$. For
\be
(g_t - g)(x) = (\square + m^2) \ 
(2\pi)^{-d/2} \! \! \int \! dp \, { \widetilde{g}(p)\over(p_0 + \omega_p)}
\, {(1 - e^{i(p_0 - \omega_p)t}) \over (p_0 - \omega_p)}  \, e^{-ipx}, 
\ee
and the expression under the integral is 
a test function because of the support properties of $\widetilde{g}$.
Making use of the decomposition $g_t = \hat{g}_t + \check{g}_t$ 
and temperateness, which implies 
\be
|| G^* ({{\check{g}_t}}^{\,*})\, \Psi^* || \leq 
\int \! dx \, |\check{g}_t (x) | \, || G^* U(x) \Psi^* ||  \leq 
\int \! dx \, |\check{g}_t (x) | \, Q (x) 
\ee
for some polynomial $Q$, one finds that   
$G^*({{\check{g}_t}}^{\, *}) \Psi^* \rightarrow 0$ as $t \rightarrow \pm
\infty$ and 
\be
G^* ({g}^*) \Psi^* = G^*({g_t}^*) \Psi^* 
= \lim_{t \,\rightarrow \,\pm \infty} G^*({{\hat{g}_t}}^{\,*})\, \Psi^*, 
\ee
strongly. Combining these facts, one gets  
\begin{eqnarray} \label{3.17}
 (A(f) \vac, G^*({g}^*) \Psi^*) & = &  
\lim_{t \,\rightarrow \,\pm \infty}
(A(\hat{f}_t) \vac, G^*({{\hat{g}_t}}^{\,*})\, \Psi^*)  \nonumber \\ 
& = & \lim_{t \,\rightarrow \,\pm \infty} \int \! dx \, {{\hat{g}_t^{\,*}}}(x)
\, (A(\hat{f}_t) \vac, G^*(x)\, \Psi^*). 
\end{eqnarray}
According to the choice of the test function $f$, its velocity 
support $\Gamma (f)$ is contained in a small neighborhood
$\Gamma_\varepsilon(\Psi_1)$ of $\Gamma (\Psi_1)$, and
consequently the operators
$A(\hat{f}_t)$ are localized in $\CO + t \, \Gamma_\varepsilon(\Psi_1)$. 
On the other hand, the operators $G^*(x)$, appearing under
the integral in (\ref{3.17}), are affiliated with $\CA(\CW_1 + x)$, 
$x \in t \, \Gamma_\varepsilon(g)$. Because of locality, they
commute with $A(\hat{f}_t)$ on their respective domains  
if $\CW_1 + t\,\Gamma_\varepsilon(g)$ is 
spacelike separated from $\CO + t \, \Gamma_\varepsilon(\Psi_1)$.

In case (a) of the statement, there holds 
$-\Gamma( \Psi_1 )  \prec  - \Gamma(g)$ and therefore
also $-\Gamma_\varepsilon( \Psi_1 )  \prec  - \Gamma_\varepsilon(g)$
if the respective neighborhoods are suitably chosen. Hence, according 
to the above geometrical considerations, the regions
$- |t| \, \Gamma_\varepsilon( \Psi_1 ) $ and  
$\CW_1 - |t| \, \Gamma_\varepsilon(g) $ are spacelike separated, and
their spatial distance increases linearly with $|t|$. 
Because of the latter fact and since $\CO$ is bounded, 
the two regions 
\pagebreak 
$\CO + t \, \Gamma_\varepsilon(\Psi_1)$ and 
$t \, \Gamma_\varepsilon(g)$ are spacelike separated if  
$t < 0$ and $|t|$ is sufficiently large. One can then 
reexpress the integral in (\ref{3.17}) according to 
\begin{eqnarray} 
\int \! dx \, {{\hat{g}_t^{\,*}}}(x) 
\, (A(\hat{f}_t) \vac, G^*(x)\, \Psi^*) & = & 
\int \! dx \, {{\hat{g}_t^{\,*}}}(x)
\, (\vac,  G^*(x)\, A(\hat{f}_t)^* \, \Psi^*) 
\nonumber \\ & = &
(G({{\hat{g}_t}})\, \vac, A(\hat{f}_t)^* \, \Psi^* ). 
\end{eqnarray}
In the latter expression, one can reverse now the passage from 
$f_t,g_t$ to their respective asymptotically dominant parts,
taking into account that, in the limit of asymptotic $t$,  
$||G(\check{g}_t) \vac|| \rightarrow 0$, 
$||A(\check{f}_t)|| \rightarrow 0$, and 
$||A({f}_t)^* \Psi^*||\leq\mbox{const}$ 
since $\Psi^*$ has compact spectral support
\cite{Bu}. Hence, by a straightforward estimate, one finds that
$(G({{\hat{g}_t}})\, \vac, A(\hat{f}_t)^* \, \Psi^* )$ 
and $(G(g_t) \, \vac, A(f_t)^* \, \Psi^*) = 
(A(f_t) \, G(g)\, \vac, \Psi^* )$ have the same limit
as $t \rightarrow - \infty$. Plugging this information into  
relation (\ref{3.17}) and making use of the asymptotic 
formula (\ref{3.9}), one arrives at 
\begin{eqnarray}
 (A(f) \vac, G^*({g}^*) \Psi^*) & = & \lim_{t \rightarrow - \infty}
 (A(f_t) \, G(g)\, \vac, \Psi^* ) \nonumber \\
 & = & ((A(f) \vac \times  G(g)\, \vac)_\inc \, , \Psi^* ).
\end{eqnarray}
In the  
resulting equation, one can replace $A(f) \vac$ by
$\Psi_1$ since $||\Psi_1 - A(f) \vac|| < \delta$, where $\delta > 0$ was
arbitrary, and 
\be
||(\Psi_1 \times G(g) \, \vac)_\inc - (A(f) \vac \times G(g)\, \vac)_\inc||
\leq \sqrt{2} \, ||\Psi_1 - A(f) \vac|| \, ||G(g) \, \vac||,
\ee
by the Fock structure of collision states. 
In view of relation (\ref{3.12}), this completes
the proof of part (a) of the statement. The proof of part (b)
is similar, but now one has to take into account that the 
regions $\CO + t \, \Gamma_\varepsilon(\Psi_1)$ and 
$t \, \Gamma_\varepsilon(g)$ are spacelike separated if  
$t > 0$ is sufficiently large. So in this case one arrives 
at an interpretation of the vectors $ G(g) \Psi_1$
in terms of outgoing collision states.
\end{proof} 
In the next step, we establish a weak form of commutation relations between
the operators $ G(x) $ and the asymptotic creation and
annihilation operators. The result will enable us to compute
scattering amplitudes.
\begin{lemma} Let $\Psi$, $\Psi^*$ be vectors with compact spectral
support in the domains of temperateness of $G$ and $G^*$,
respectively, and let $f,g$ be test functions whose 
Fourier transforms have support in small neighborhoods of 
points on the mass shell. Then 
\begin{statement}
\item (a) \ $(G(g) \, \Psi, \, {A(f)_\inc}^\bullet \, \Psi^*) = 
({A(f)_\inc}^{\bullet \, *} \, \Psi, \, G^*(g^*) \, \Psi^*) $
\ \ if \ \ $\Gamma(g) \prec \Gamma(f)$,
\item (b) \ $(G(g) \, \Psi, \, {A(f)_\out}^\bullet \, \Psi^*) = 
({A(f)_\out}^{\bullet \, *} \, \Psi, \, G^*(g^*) \, \Psi^*) $
\ \ if \ \ $\Gamma(f) \prec \Gamma(g)$.
\end{statement}
Here the symbol $X^\bullet$ stands for both, the operator $X$ and its
adjoint $X^*$. \label{L3}
\end{lemma}
\begin{proof} The argument is very similar to the proof of 
the preceding lemma and it therefore suffices to indicate the main steps.
In case (a) one has, in view of
the fact that $G(g_t) \, \Psi = G(g) \, \Psi$, $t \in \RR$, and 
the asymptotic relation (\ref{3.8}), 
\begin{eqnarray}
 (G(g) \, \Psi, \, {A(f)_\inc}^\bullet \, \Psi^*) & = & 
 (G(g_t) \, \Psi, \, {A(f)_\inc}^\bullet \, \Psi^*) \nonumber \\
 = \lim_{t \rightarrow - \infty}  
 (G(g_t) \, \Psi, \, A(f_t)^\bullet \, \Psi^*) 
& = & \lim_{t \rightarrow - \infty}  
 (G(\hat{g}_t) \, \Psi, \, A(\hat{f}_t)^\bullet \, \Psi^*), 
\end{eqnarray}
where, in the last step, $f_t, g_t$ have been replaced  
by their asymptotically
dominant parts. Since $\Gamma(g) \prec \Gamma(g)$,
the regions $\CW_1 + t \, \Gamma_\varepsilon(g)$ and 
$\CO + t \, \Gamma_\varepsilon(f)$ are spacelike separated for 
$t < 0$ and $|t|$ sufficiently large. Hence, by locality, 
\begin{eqnarray} 
\lim_{t \rightarrow - \infty}  
 (G(\hat{g}_t) \, \Psi, \, A(\hat{f}_t)^\bullet \, \Psi^*) & = &
\lim_{t \rightarrow - \infty}  
 (A(\hat{f}_t)^{\bullet *} \, \Psi, \, G^*({{\hat{g}_t}}^{\,*}) \, \Psi^*)
\nonumber \\
= \lim_{t \rightarrow - \infty}  
 (A(f_t)^{\bullet *} \, \Psi, \, G^*({g_t}^*) \, \Psi^*) & = &
 ({A(f)_\inc}^{\bullet \, *} \, \Psi, \, G^*(g^*) \, \Psi^*),
\end{eqnarray} 
where, in the second equality, the transition from 
$f_t, g_t$ to the  asymptotically dominant parts has been reversed 
and, in the last step, the asymptotic relation (\ref{3.8})
has been used as well as the fact that 
$G^*({g_t}^*) \, \Psi^* = G^*(g^*) \, \Psi^*$, $t \in \RR$. 
This establishes statement (a). The 
proof of (b) is analogous.
\end{proof}
In a final step,  we have to determine the 
spectral support of $G\vac$ with respect to the spatial 
momentum operators $\bP$ in order to see which single particle 
states can be generated by $G$ from the vacuum. 
\begin{lemma} Let $G$ be a polarization--free generator which 
is affiliated with the algebra $\CA(\CW_1)$. The spectral support of 
$G\vac$ with respect to the spatial momentum operators 
$\bP = (P_1,P_2, \dots P_{d-1})$ is equal to $\RR\times\CC$, 
where $\CC \subset \RR^{d-2}$ is a closed set with open 
interior. \label{L4}
\end{lemma}
\begin{proof}
Let $A$ be any local operator which is localized in ${\CW_1}^\prime$.
Since $x \rightarrow G(x)\vac$ and $x \rightarrow G^*(x)\vac$ are 
solutions of the Klein--Gordon equation of mass $m$, the commutator
function 
\be
x \rightarrow C(x) = (A\vac, G(x) \vac) - 
(G^*(x) \vac, A^* \vac)
\ee
can be represented in the form
\be \label{3.24}
C(x) = \int \! {d\bp \over 2 \omega_p} \, 
(K_+(\bp) \, e^{i\omega_p x_0} - K_-(\bp) \, e^{-i\omega_p x_0}) 
e^{-i \bfp \bfx}.
\ee
Here the functions $K_\pm$ are given by
\begin{eqnarray} \label{3.25}
K_+(\bp) & = & {(A\vac)(\bp)}^* (G\vac)(\bp), \nonumber \\
K_-(\bp)  & = & (G^*\vac)(-\bp)^* (A^*\vac)(-\bp), 
\end{eqnarray}
where $\bp \rightarrow (A\vac)(\bp)$ is the momentum space wave 
function of the single particle vector $\Em A\vac$, and similarly
for the other terms. 

Because of the localization properties of the operators
$A$ and $G$, the commutator function $x \rightarrow C(x)$
and its time derivative vanish at time $x_0 = 0$ in the 
half space $\{\bx \in \RR^{d-1}: x_1 > 0 \}$. In view of 
the representation (\ref{3.24}), this implies that the 
functions 
\be
\bp \rightarrow {1\over\omega_p} (K_+(\bp) - K_-(\bp)) 
\quad \mbox{and} \quad \bp \rightarrow (K_+(\bp) + K_-(\bp))
\ee 
can be analytically continued in $p_1$ into the 
lower half plane. As $\bp \rightarrow \omega_p$
is analytic in $p_1$ in a strip about the origin,
the functions $\bp \rightarrow K_\pm(\bp)$ 
can likewise be analytically continued in $p_1$ into some  
strip of the lower half plane. 

Now if $\CU \subset \RR$ and $\CV \subset \RR^{d-2}$ are
open sets such that $\bp \rightarrow (G\vac)(\bp)$ 
vanishes for almost all spatial momenta 
$\bp = (p_1,\bp_\perp)$ with  
$p_1 \in \CU$ and $\bp_\perp \in \CV$,
the function $\bp \rightarrow K_+(\bp)$ vanishes for these momenta 
as well. Being the boundary value of an analytic
function with respect to $p_1$, $K_+(\bp)$ therefore vanishes for all 
$p_1 \in \RR$ and $\bp_\perp \in \CV$. Since $A \in \CA({\CW_1}^\prime)$
was arbitrary and the set of 
single particle states $\Em A \vac$, $A \in \CA({\CW_1}^\prime)$,
is dense in $\Em \CH$, this implies $(G\vac)(\bp)=0$ for $p_1 \in \RR$ and 
$\bp_\perp \in \CV$. Thus the complement of the support of 
$\bp \rightarrow (G\vac)(\bp)$ in momentum space $\RR^{d-1}$
has the form $\RR\times\CV$, where $\CV \subset \RR^{d-2}$ is
open. Hence, disregarding sets of measure $0$, the statement 
follows since $G\vac$ is different from zero.
\end{proof}
We have accumulated now sufficient information in order to proceed to 
the computation of the scattering amplitudes in the underlying 
theory, provided the dimension of spacetime is larger than two. Let 
$\bp = (p_1,\bp_\perp)$ be any vector 
in the spectral support of $G\vac$ with respect to
the spatial momentum operators and let $p_1 < 0$ and 
$\bp_\perp \neq 0$. We  
pick a test function $g$ whose Fourier transform has support in 
a sufficiently small neighborhood of $(\omega_p,\bp)$,
and a single particle state $\Psi_1$ which is an element of the 
domain of temperateness of $G$ with spectral support in a small
neighborhood of  $(\omega_p,-\bp)$. Hence 
$\Gamma(g) \prec \Gamma(\Psi_1)$ and consequently 
$G (g) \Psi_1 = {(G(g) \vac \times \Psi_1 )}_\inc $ by Lemma \ref{L2}.

As $\bp_\perp \neq 0$,
there are spatial momenta $\bq$ with $q_1 < p_1$ and $|\bq| = |\bp|$
(here the dimension of spacetime enters). For any such $\bq$,
we choose a test function $f$ whose Fourier transform has support
in a small 
neighborhood of $(\omega_q,\bq)$ such that $\Gamma(f) \prec \Gamma(g)$.
Finally, we pick a single particle state $\Psi_1^*$ in the domain
of temperateness of $G^*$ with spectral support about
$(\omega_q,-\bq)$. After these preparations, we can apply 
Lemma \ref{L3} and compute 
\begin{eqnarray}
& ((G(g) \vac \times \Psi_1 )_\inc, (A(f)\vac \times \Psi_1^*)_\out) =
(G(g)\Psi_1, A(f)_\out \Psi_1^*) & \nonumber \\
& = (A(f)_\out^* \Psi_1, G^*(g^*)\Psi_1^*) = 0, & 
\end{eqnarray}
where, in the last step, we used the fact that 
${A(f)_\out}^* \Psi_1 = (A(f) \vac, \Psi_1)\,\vac = 0$
since $\Gamma(f) \prec \Gamma(\Psi_1)$. 
Varying $f,g$ and $\Psi_1, \Psi_1^*$ 
within the above limitations, it follows that 
elastic scattering processes of two particles with initial momenta about 
$\bp, -\bp$ and final momenta about $\bq, -\bq$ -- although
admitted by the energy--momentum conservation law -- do not 
occur in the underlying theory. 

This result implies  
that the elastic two--particle scattering amplitude $T$ vanishes 
identically. For the proof of this statement, we recall 
that in a relativistic theory $T = T(s,t)$ is a 
distribution with respect to the invariants $s,t$
(the squares of the energy in the center of mass system
and the momentum transfer, respectively). Thus for 
$\bp,\bq$ as above, $s = 4(m^2 + \bp^2)$ 
and $t = - 2 \bp^2 (1 - \cos \theta)$, where 
$\theta$ is the scattering angle. It is a well--known 
consequence of locality, relativistic covariance,
and the form of the energy--momentum spectrum 
that $T(s,t)$ is, in the physical region 
$s \geq 4m^2 - t$ for fixed $t \leq 0$, the boundary value 
from $\mbox{Im} s > 0$ of an analytic
function in the cut $s$--plane. On the other hand, for fixed $s$, it is 
analytic in the variable $\cos(\theta)$ in the  
Lehmann ellipse \cite{Le} with foci at $\pm 1$ and semi-minor 
axis of length $6m^2/\sqrt{s(s-4m^2)}$, 
cf.\ \cite{Ma} for an exposition of these
basic facts. 

By the preceding computations, we know that the scattering
amplitude vanishes for $s,t$ in some open set. 
In fact, taking into account that
all momenta $\bp = (p_1,\bp_\perp )$ with $p_1 \in \RR_-$
belong to the spectral support of $G\vac$ for some fixed
$\bp_\perp \neq 0 $ (cf.\ Lemma \ref{L4}) 
and varying $\bq$ within the above limitations, we get 
\be
T(s,t) = 0 \quad \mbox{for} \quad s > 4(m^2 + |\bp_\perp|^2), \ \
0 > t > -4|\bp_\perp|^2.
\ee
By analyticity in $\cos(\theta)$, this equality extends to
all scattering angles and hence to all $t$ for the given
range of $s$. Analyticity of $T(s,t)$ in $s$
then implies that the scattering amplitude vanishes everywhere. 
It is a well--known consequence of this result that then there can be 
no non--trivial multi particle scattering or particle
production either \cite{Ak}. These implications hold in any 
number $d > 2$ of spacetime dimensions \cite{Br}.
So we arrive at the following statement.
\begin{theorem} If in a local, relativistic quantum field theory
of a scalar massive particle 
in $d > 2$ spacetime dimensions there exists a {temperate} 
polarization--free generator, then 
the scattering matrix is trivial.
\end{theorem}
For the sake of simplicity, we have restricted attention
in the preceding analysis to theories describing a  
single scalar massive particle. But
it should be clear from our discussion that similar results 
hold in theories with a more complex particle spectrum. 
There one finds that particles whose states can be 
generated from the vacuum
by temperate polarization--free generators do not participate in collision
processes. For the derivation of this result it is actually not necessary 
to assume that the collision states can be constructed 
by local operators. Cone--like localized ``interpolating 
operators'', whose existence has been established in all theories of massive 
particles \cite{BuFr}, are completely sufficient for
the proof. We therefore conclude that in the presence 
of interaction there is no room for 
temperate polarization--free generators in more than two
spacetime dimensions. 
\vspace*{1mm} 
\section{\hspace{-2mm}Polarization--free generators in two dimensions}
\setcounter{equation}{0}

The analysis in the preceding section did not lead to any restrictions
on the form of the elastic scattering amplitudes in two spacetime 
dimensions. For configurations of asymptotic particle momenta which
would allow one to show that the scattering amplitudes have to vanish   
in the presence of temperate polarization--free generators 
cannot occur in this case because of the energy--momentum 
conservation law. So non--trivial theories admitting
polarization--free generators can and do exist in two spacetime 
dimensions \cite{Sch}. It seems therefore worthwhile to have a closer
look at the type of constraints imposed on such theories from 
the present general point of view. 

In order to abbreviate this discussion, we assume in the following 
that the domains of temperateness of $G$ and $G^*$ contain incoming
and outgoing collision states for arbitrary configurations of 
particle momenta. Lemma \ref{L3} then implies that on these states
\begin{eqnarray}
G(g) \, {A(f)_\inc}^* =  {A(f)_\inc}^*  \, G(g) 
& \ \mbox{if} \ &  \Gamma(g) \prec \Gamma(f),  \label{4.1} \\
G(g) \, {A(f)_\out} =  {A(f)_\out} \ G(g) 
& \ \mbox{if} \ &  \Gamma(f) \prec \Gamma(g), \label{4.2}
\end{eqnarray}
and similarly for $G^*(g^*)$. For the lemma says that 
on the domain of temperateness 
$G^*(g^*)^* {A(f)_\inc}^* \supset {A(f)_\inc}^* G(g)$
if $ \Gamma(g) \prec \Gamma(f)$, say, and with the above domain 
assumptions one can replace the triple--star expression 
$G^*(g^*)^*$ by $G(g)$. 

It turns out that these commutation relations imply that there 
can be no particle production in the underlying theory. In the proof
of this statement, we make use of the following lemma.
\begin{lemma} Let $f,g_1,\dots g_n$ be test functions whose 
Fourier transforms have \mbox{support} about points on the 
mass shell such that 
$\Gamma(g_1) \prec \dots \prec \Gamma(g_n) \prec \Gamma (f)$
and let $A,A_1,\dots A_n \in \AO$ be arbitrary local operators.
Then 
\be
{A(f)_\inc}^* \, (A_1(g_1)\vac \times \cdots \times A_n(g_n)\vac)_\out = 0.
\ee
\end{lemma}
\begin{proof} The proof is based on induction in $n$. For $n=1$, 
one has 
\be
{A(f)_\inc}^* \, A_1(g_1)\vac = 
(A(f) \vac, A_1(g_1)\vac) \, \vac = 0
\ee
because of the support properties of $f, g_1$ in momentum space.
Assuming that the statement holds for $n$, let $g$ be a test
function whose Fourier transform has support about
points on the mass shell such that 
$\Gamma(g_1) \prec \dots \prec \Gamma(g_n) 
\prec \Gamma(g) \prec \Gamma (f) $. It then follows from 
relation (\ref{4.2}) that 
\begin{eqnarray}
& G(g) \, A_1(g_1)_\out \cdots A_n(g_n)_\out \, \vac  = 
 A_1(g_1)_\out \cdots A_n(g_n)_\out \, G(g) \vac & \nonumber \\
& = (A_1(g_1)\vac \times \cdots \times A_n(g_n)\vac \times G(g) \vac)_\out. &
\end{eqnarray}
Hence, by relation (\ref{4.1}) and the induction hypothesis, one
obtains 
\begin{eqnarray} \label{4.6}
\lefteqn{\hspace{-2em} {A(f)_\inc}^* \,  (A_1(g_1)\vac \times \cdots 
\times A_n(g_n)\vac \times G(g) \vac)_\out}  \nonumber \\
& = & {A(f)_\inc}^* \, G(g) \, 
A_1(g_1)_\out \cdots A_n(g_n)_\out \, \vac  \nonumber \\
& = & G(g) \, {A(f)_\inc}^* \, A_1(g_1)_\out \cdots A_n(g_n)_\out \, \vac 
= 0. 
\end{eqnarray}
Now let $A_{n+1} \in \AO$ be any local operator and 
$g_{n+1}$ any test function
such that $\Gamma(g_1) \prec \dots \prec \Gamma(g_n) 
\prec \Gamma(g_{n+1}) \prec \Gamma (f) $. There exists for given  
$\delta > 0$ a test function $g$ as in the preceding step such
that $||A_{n+1}(g_{n+1})\,\vac - G(g)\,\vac||<\delta$. For the spectral
support of $G\vac$ consists of the whole mass shell according to
Lemma~\ref{L4} and consequently the set of vectors 
$\{ G(g) \, \vac : \supp \, \widetilde{g} \subset \Delta \} $ is,
for any compact set $\Delta \subset \RR^2$, 
dense in the corresponding spectral subspace $E(\Delta) \, E_m \CH$ of
single particle states. As the collision states are continuous
with respect to their single particle components, one can thus replace 
in equation (\ref{4.6}) the vector $G(g) \vac$ by 
$A_{n+1}(g_{n+1})\vac$, proving the statement. 
\end{proof}

\vspace*{1mm}
Let us consider now an incoming collision state of two particles
with momenta $p_1, p_2$ on the mass shell and an outgoing state
of $n>2$ particles with mutually different momenta $q_1,\dots q_n$.
Taking advantage of the fact that in $d=2$ spacetime dimensions
the momenta on the mass shell are linearly ordered, we may assume
without loss of generality that $q_1 < \cdots < q_n$. Now if the 
incoming state is to evolve with non--zero probability into this 
outgoing state, the energy--momentum conservation law requires that 
$p_1 + p_2 = q_1 + \cdots + q_n$. Taking into account that the
linear order of momenta is preserved under proper othochronous 
Lorentz transformations, it is not difficult to see that at least one
of the incoming particle momenta, say $p_1$, has to be strictly 
larger than any one of the outgoing momenta, i.e.\
$q_1 < \cdots < q_n < p_1$. We pick now test functions $f_1,f_2$ and
$g_1,\dots g_n$ which, in momentum space, have support about
$p_1, p_2$ and $q_1,\dots q_n$, respectively, such that 
$\Gamma(g_1) \prec \dots \prec \Gamma(g_n) \prec \Gamma (f_1)$.
Thus, for any choice of local operators 
$A_1, \dots A_{n+2} \in \AO$, we obtain with the help of the preceding
lemma
\begin{eqnarray}
\lefteqn{\big( (A_1(f_1)\vac \times A_2(f_2)\vac)_\inc , \, 
(A_3(g_1)\vac \times \cdots \times A_{n+2}(g_n)\vac)_\out \big) } \nonumber \\
& & = (A_1(f_1)_\inc \, A_2(f_2) \vac, \, 
(A_3(g_1)\vac \times \cdots \times A_{n+2}(g_n)\vac)_\out) \nonumber \\
& & = (A_2(f_2) \vac, \, {A_1(f_1)_\inc}^* \,
(A_3(g_1)\vac \times \cdots \times A_{n+2}(g_n)\vac)_\out) = 0.
\end{eqnarray}
But the set of collision states with non--overlapping momenta
is dense in the set of all collision states, so we arrive
at the conclusion that an 
incoming collision state of two particles can never evolve into an 
outgoing collision state containing more than two particles, i.e.\ there is no
particle production in the underlying theory.
\begin{theorem} 
If a local, relativistic quantum field theory in $d=2$ spacetime
dimensions admits temperate polarization--free generators, there is 
no particle production.
\end{theorem}

This result shows that in $d=2$  dimensions temperate
polarization--free generators can only exist in the presence of 
additional conservation laws, besides energy--momentum conservation.
We have illustrated this fact on the example of the particle 
number. By a more refined analysis, one can show that also the
individual particle momenta have to be preserved in multi--particle
collisions. This brings us close to the structure of scattering 
amplitudes found in completely integrable models. In particular, the  
apparently general Ansatz for correlation functions of polarization--free 
generators, proposed by one of the authors in \cite{Sch,Sch1}, falls back
to this special class of theories. Moreover, there are 
indications in the present general setting that temperate 
polarization--free generators necessarily have algebraic properties
of the type found in these examples. Thus the notion of temperate 
polarization--free generator may not only be useful for the
characterization of such integrable models, but it might also serve 
as a tool for their general analysis and classification. This
interesting aspect of the present investigation will be discussed
elsewhere. 
\section{\hspace{-2mm}Concluding remarks}
\setcounter{equation}{0}
Harry Lehmann, one of the pioneers of the rigorous approach to 
relativistic quantum field theory, liked to mock at the sometimes 
cumbersome subtleties appearing in this setting as ``problems 
of inessential selfadjointness''. But his scientific work provides
ample evidence to the effect that he was willing to invest 
mathematical diligence and care where the physical context required it. 

In the present article, we have encountered a surprisingly 
subtle feature of relativistic quantum field theory: Mathematics 
tells us, on one hand, that any such theory accommodates 
well--defined polarization--free generators. Physics, on the 
other hand, implies that these generators necessarily  
have rather peculiar domain properties which do not allow one 
to apply methods of Fourier analysis. Their 
relation to the asymptotic particle interpretation is 
thereby obscured. Being sloppy with regard to these domain 
properties, one would be led to the unpleasant conclusion that the 
fundamental postulates of relativistic quantum field theory 
exclude interaction in more than two spacetime dimensions.
Thus it is this subtle point which provides the 
loophole for theories with non--trivial interaction.

Temperate polarization--free generators exist, however, in
two--dimensional integrable models and the present results 
indicate that they are a distinctive
feature of such theories. This fact may be attributed 
to the presence of large groups of conservation laws in such 
theories which help to restrain the polarization effects of local 
operations. We believe that a more detailed investigation
of these temperate generators is warranted and will lead to a better  
understanding of the specific features of these models.
 
There is another aspect of the present analysis which 
deserves mentioning, namely the problem of particle statistics 
in low dimensions. We have discussed here only the simple case of 
bosons, the case of fermions being similar. But it is well--known 
that particles in two and three spacetime 
dimensions can also have anyonic or plektonic
statistics (cf.\ \cite{FrReSch,FrMa} for a systematic analysis of this 
issue). There is also a general collision theory for such 
particles \cite{FrGa}, but it is an open question whether 
there exists some kind of associated free fields. 

A negative result 
to that effect is due to Mund \cite{Mu}, who proved that there are 
no operator--valued solutions of the Klein--Gordon equation which 
generate such particles from the vacuum and are localized in salient 
(pointed) spacelike cones. This {\it ad hoc} assumption about the 
localization is, however, crucial for the proof of this 
no--go theorem. In fact, as in theories of massive anyons and plektons there 
are still cone--like localized (vacuum polarizing) operators 
which generate the states of physical interest from the vacuum, 
one can establish the existence of \mbox{wedge--localized} 
polarization--free generators 
for these particles. Taking 
temperateness as an additional input, 
it may well be possible to construct from these generators 
in a systematic manner examples of anyonic or even plektonic 
theories which come close to the idea of a free field theory. 

\end{document}